\documentclass[12pt, aaspp4, flushrt, mathptmx]{aastex}

\usepackage{epsf}

\newcommand{\comments}[1]{}

\begin{document}

\sffamily

\title{Small Scale Field Emergence and Its Impact on Photospheric Granulation}

\author{ V. Yurchyshyn, K. Ahn, V. Abramenko, P. Goode, W. Cao}

\affil{\it Big Bear Solar Observatory, New Jersey Institute of  Technology, Big Bear City, CA 92314, USA}

\begin{abstract}
We used photospheric intensity images and magnetic field measurements from the New Solar Telescope in Big Bear and Helioseismic Magnetic Imager on board Solar Dynamics Observatory (SDO) to study the the effect that the new small-scale emerging flux induces on solar granulation. We report that emerging flux appears to leave different types of footprint on solar granulation: i) diffuse irregular patches of increased brightness, ii) well defined filament-like structures and accompanied bright points, and iii) bright point-like features that appear inside granules. We suggest that the type of the footprint depends on the intensity of emerging fields. Stronger fields, emerging as a part of large magnetic structure, create on the solar surface a well defined filamentary pattern with bright points at the ends of the filaments, while weak turbulent fields are associated with bright patches inside the host granule.
\end{abstract}

\section{Introduction}

New magnetic fields constantly emerge into the solar atmosphere at variety of spatial scales ranging from hundreds of megameters (active regions) to a fraction of a megameter (granule). Each emergence episode injects energy into the solar atmosphere, that further cascades along the spatial scales to be later released in form of solar flares, coronal mass ejections, coronal heating, plasma flows, etc. It is essential to understand the detail of the flux emergence process, especially the interaction of the new injected flux with the pre-existing fields as well as the interaction of emerging fields with convection flows and their effect the solar photosphere.

Strong large scale emerging fields significantly modify solar granulation. Granules may become elongated \citep{ishikawa_tsuneta_2010}, transient darkening and new bright points often appear at the emergence site \citep{Cheung_2007}. However, it is less known about the footprint of weak small-scale ($<$ 2~Mm) emerging fields. Small-scale and weak ($<$ 1~kG), predominantly horizontal fields were first reported by \cite{Lites_1996}. They are short-lived, typically lasting 5 minutes or less.  Later observations and theoretical considerations \citep{Bart_2002, Khomenko_2003, Dominguez_2003, Rachkovsky_2005, centeno2007, ishikawa_2008,lites_2008, orozco, martinez2009,Gomory_2010, Zhou_2010,ishikawa_tsuneta_2010,Ishikawa_2011} expanded the earlier studies and currently known properties of small-scale fields could be summarized as follows. The fields emerge at scales less than  1-2~Mm \citep[\textit{e.g.},][]{Bart_2002, Lites_1996,centeno2007,orozco}. Their lifetime ranges from 1 to 20~min and seems to be correlated to the amount of emerging flux \citep[\textit{e.g.},][]{Zhou_2010}, which is of order of 10$^{17}$ Mx \citep[\textit{e.g.},][]{Khomenko_2003, Gomory_2010}. The small-scale magnetic elements have field strength less than 1~kG \citep[\textit{e.g.},][]{Lites_1996} and may cover nearly 60\% of the area, most of which corresponds to intergranular lanes \citep{Dominguez_2003}. Significant fraction of the magnetic flux appears as $\Omega$ loops, which emerge at the rate of 0.02 loops per hour and arcsec$^2$ \citep{martinez2009}, bringing to the solar surface up to 10$^{24}$ Mx of flux daily  \citep{Lites_1996}. Only about 23\% of the loops are observed to reach the chromosphere \citep{martinez2009}. While some authors conclude that the emerging fields do not seem to be correlated with any brightness structures and/or affect neighbouring G-band bright points \citep{Gomory_2010}, others point out that emergence disturbs the granulation and is associated with granule extension \citep{ishikawa_2008}.

In this paper we will focus on the footprint of small-scale emerging fields on the granulation field at the site of emergence.

\section{Observations}

\noindent In this study we used three different data sets that include photospheric images and magnetic field measurements from the New Solar Telescope \citep[NST,][]{goode_nst_2010, goode_apjl_2010} as well as Stokes data from SDO's Helioseismic and Magnetic Imager \citep[HMI,][]{hmi,hmi2}.

The NST  intensity images were acquired using the broad-band filter imager (BFI) equipped with a TiO 1.0~nm interference filter centered at 705.7~nm. The pixel scale and field of view (FOV) for the BFI were 0.0375'' and 77''. All data were acquired with the temporal sampling of about 10~s and with the aid of an adaptive optics system that utilizes a 97 actuator deformable mirror. The time series were speckle reconstructed employing the KISIP speckle reconstruction code \citep{kisip_code} and then rigidly co-aligned using cross-correlation and de-stretched to remove distortions due to residual seeing effects of the Earth's atmosphere.

To detect magnetic field emergence we used 1" resolution 45 s cadence full-disk SDO/HMI line-of-sight HMI magnetograms and 12~min averaged HMI Q and U maps available at JSOC\footnote{Joint Science Operations Center, {\tt http://jsoc.stanford.edu/ajax/exportdata.html}} and polarization maps from NST's InfraRed Imaging Magnetograph \cite[IRIM,][]{Cao_IRIM}. The IRIM is an imaging solar spectro-polarimeter that uses a pair of Zeeman sensitive Fe I lines present in the near infrared at 1564.85~nm and 1565.29~nm. It is based on a 2.5~nm interference filter, a 0.25~nm birefringent Lyot filter, and a Fabry-P{\'e}rot etalon and it is capable of providing a bandpass as low as 0.01~nm over a FOV of 50''. With the aid of adaptive optics, we obtained circular and linear polarization images with spatial and temporal resolution of 0".2 and 43 s, respectively.

NST/TIO and IRIM data from Aug 12, 2011 were used to explore field emergence within a granule (Sec. \ref{diffuse}), NST/TiO and SDO/HMI data for Aug 30, 2010 were used to study details of filamentary field emergence (Sec. \ref{filament}), while NST/TiO images from Sept 27, 2010 were used to study bright point formation inside a granule (Sec. \ref{filament}). Aug 30, 2010 data were obtained while the NST was pointed at active region NOAA 11102 located at 20W27N of the heliospheric coordinate system, while other two data sets were collected in quiet sun areas.

\section{Results}

\subsection{Case of small scale diffuse flux emergence}
\label{diffuse}

\begin{figure}[t]
\centerline{\epsfxsize=6.5truein  \epsffile{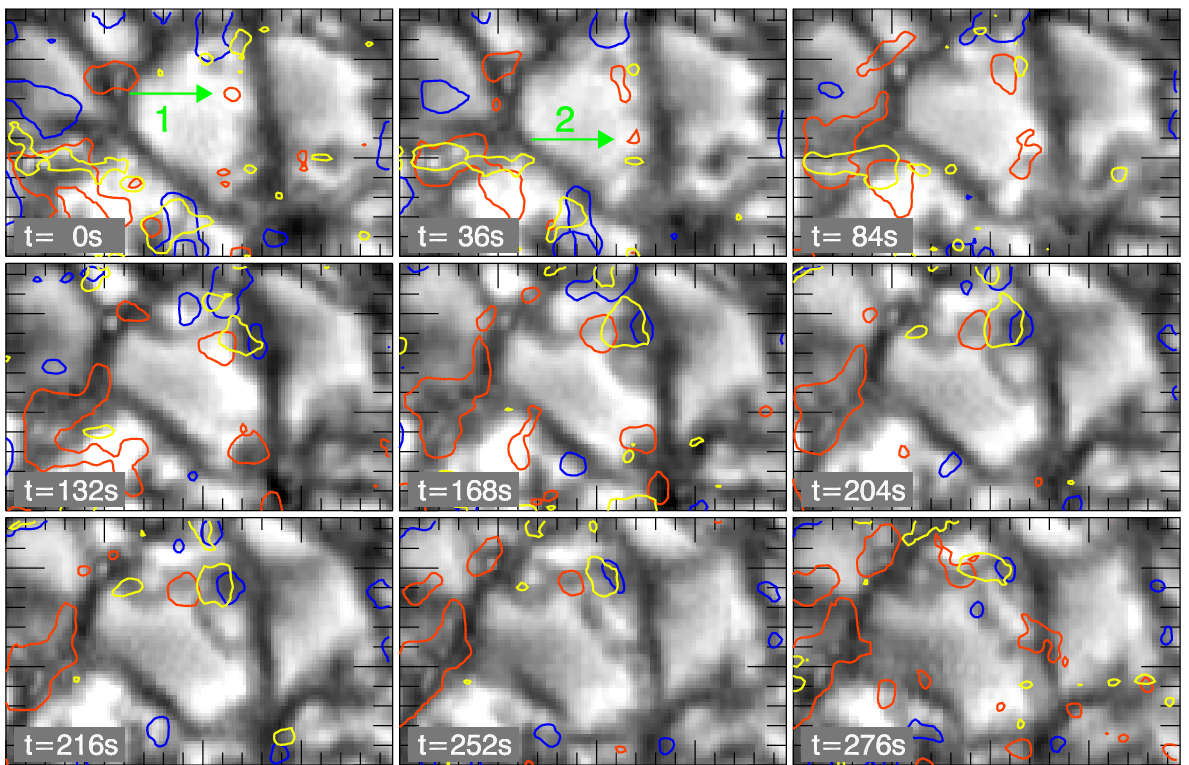}}
\caption{Sequence of TiO images showing evolution of a granule. The over-plotted contours show magnetic field components as measured with the IRIM instrument. The red (blue) contours represent negative (positive) circular polarisation signal and the yellow contours are linear polarization. The contours are plotted at 0.002 $I/I_c$ level. Time intervals of each image is shown in the lower left corner. The appearance of new flux is accompanied by development of a bright patch inside the host granule. Short tick marks separate 0.2~Mm spatial intervals.}
\label{event5_vt}
\end{figure}

\noindent Figure \ref{event5_vt} shows several instances of new flux appearance inside granules associated with bright granular substructures. One flux emergence event occurred near the upper part of the granule (arrow 1 in t=0~s panel) and it was co-spatial with a bright intensity structure. This event was developing differently from the typical sequence, when horizontal fields appear first and the vertical fields follow. In our case, circular polarization (negative polarity, red contour) appeared first followed by horizontal fields (t=84~s, yellow contour).  Approximately 2 min after the emergence began (t=132~s), a positive polarity patch appeared. This emerging magnetic configuration (see \textit{e.g}., t=132~s frame) is consistent with the idea of an emerging $\Omega$ loop.

Another flux appearance (arrow 2 in t=36~s panel) occurred nearly simultaneously with the previous. This second event seems to be similarly co-spatial with an intensity structure. Comparing t=0~s and t=84~s frames one may conclude that the bright substructure was correlated in space and time with the intensification of the corresponding magnetic patch. However, unlike the first event, no associated positive polarity appeared in close proximity to the red contour. A weak linear polarization signal appeared only in one frame and it could have been the result of data noise. This event was short lived and had already disappeared in t=204~s frame. The first event was still seen in the last frame, which was the last magnetogram of our data set. The two associated intensity features were evolving along with the granule and by t=132~s they formed a bright granular lane thought to be associated with horizontal vortex tubes \citep{Steiner_2010,Yurchyshyn_2011}. This bright granular lane showed signs of fading and disappeared by t=276~s.

There are many more cases when the magnetic and bright granular patches are correlated in time and space. Usually a bright substructure appear at the very early stage of emergence before the magnetic signal is detected and often fade from the FOV before the magnetic signal does. Sometimes the bright patches are observed as bright granular lanes, however most of the time they rather are small and of irregular shape with typical size of 0.2-0.6~Mm. Another important point to mention is that the newly appearing flux does not necessarily always have a three-component configuration (positive, negative, and  horizontal fields) consistent with the emerging of an $\Omega$ loop: more often than not, any possible combination of these fields may appear on the solar surface. However, the common feature in all the cases is the development of a diffuse intensity structure within the host granule.

\subsection{Case of small scale filamentary flux emergence}
\label{filament}

\begin{figure}[t]
\centerline{\epsfxsize=5truein  \epsffile{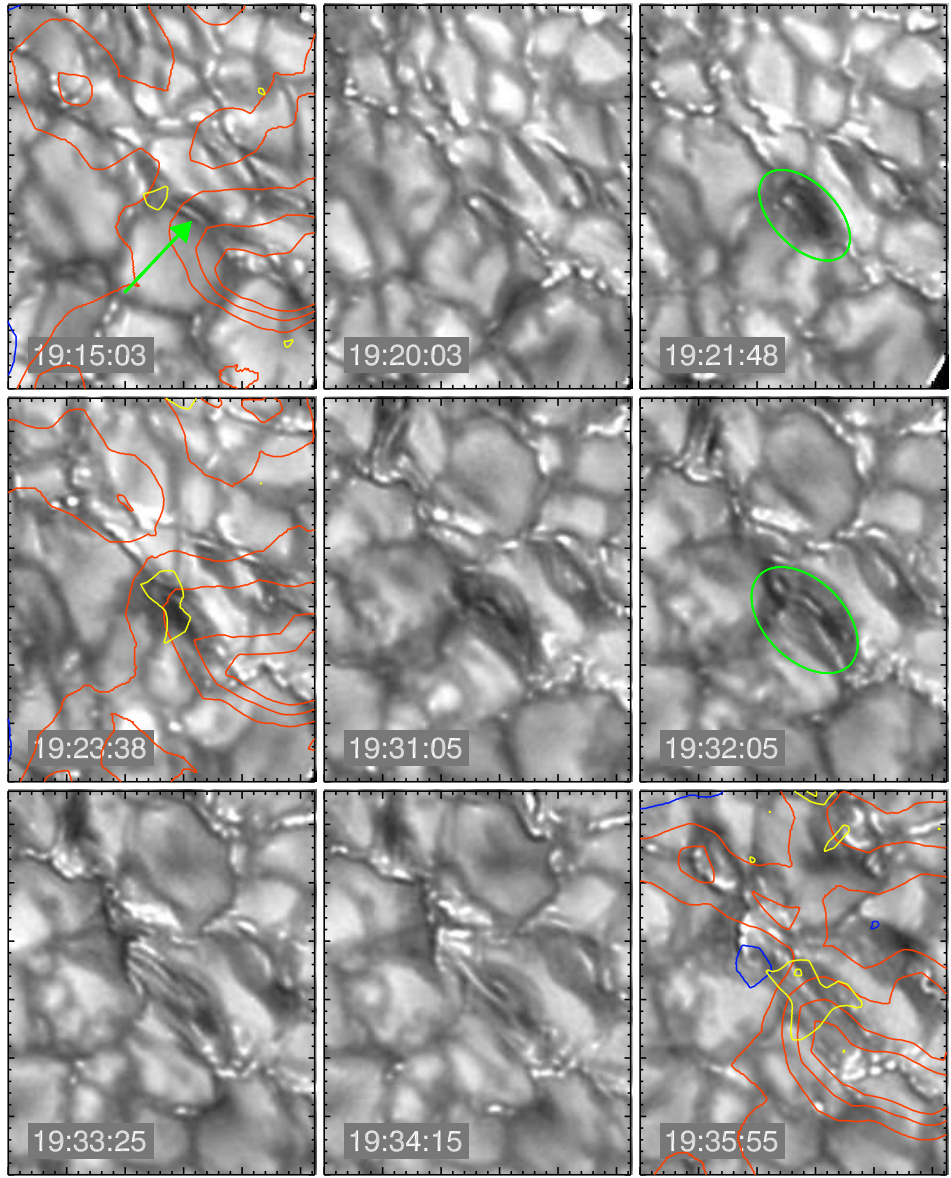}}
\caption{Sequence of TiO images showing response of a granular field to flux emergence. The over-plotted red (blue) contours show SDO/HMI negative (positive) line-of-sight flux density and the yellow contours are total SDO/HMI linear polarization. The red contours are plotted at 50, 100, 150, and 200~G levels. The blue contour is the polarity inversion outlining weak positive fields. The yellow contours are plotted at 100 and 140~g levels.The appearance of new flux is accompanied by development of a narrow bright filament(s) and a transient darkening surrounding the filament. Short tick marks separate 0.2~Mm spatial intervals.}
\label{aug30}
\end{figure}

\noindent In Figure \ref{aug30} we present a flux emergence event that produced a significant footprint on the solar photosphere. The background of each panel are series of NST/TiO images taken on August 30, 2010 in  NOAA AR 11102. Three panels 12~min apart are overplotted with contours of HMI magnetic field measurements: the red contours are plotted at -50, -100, -150, and -200~G levels of the line-of-sight field, the blue contour is the polarity inversion line and the yellow contours are plotted at 100 and 140~G levels of the horizontal field.

The object of our interest was the small filament-like feature developing in the granulation field (arrow in 19:15:03 panel) between two adjacent granules. It was well defined and had a high contrast bright core and much dimmer surroundings. It was approximately 1~Mm long and appeared in the midst of negative polarity fields (red contours). A small associated horizontal field patch (HFP, yellow) was detected in the early stage of emergence and is present in the two following $\sqrt(Q^2+U^2)$ maps each taken 12 min apart. The HFP was growing as the TiO feature was developing and increasing in size. The well pronounced correlation between the TiO and HFP indicates that the latter is a real magnetic structure rather than measurement noise.

The TiO feature was developing rapidly and going through significant changes. In the very early stage a bright point appeared at one end of it (19:20:03 panel). Immediately after the entire feature appeared to be situated inside a transient darkening, possibly indicating significant plasma down flows \citep{Cheung_2007}. The darkening became less intense by approx. 19:27~UT, although weak dark lanes were encircling the filaments nearly during the entire life time of the event. Several new aligned filaments appeared later in the TiO images, as if they were individual threads of a larger bundle of magnetic flux tubes. The TiO data did not produce any hints of possible fine structure in these filaments except the brighter central core seen in each thread. As the system evolved, the filaments became longer and their ends were shifting in opposite directions, consistent with the picture of merging of $\Omega$ type loops.

At the time when the TiO filamentary structure showed signs of fading, the HMI instrument detected a small patch of positive polarity fields, which appeared in an area previously free of any strong fields and were co-spatial with one of the footpoints of the emerging feature. The associated polarization signal was small, so that we only plot the polarity inversion line with a blue contour. The appearance of this weak positive polarity suggests that the other (stronger, lower right) footpoint had the same polarity as the surrounding area (negative, red). The negative footpoint displayed ``purposeful'' movement toward the cluster of like polarity bright points, leading us to speculate that the emerging field was connected to a larger magnetic structure. The entire event lasted for nearly 20~min, after which the filament-like features disappeared and normal granulation field was restored in the area. The final extent of the event was comparable to the size of two granules (i.e., about 3.5~Mm).

\begin{figure}[t]
\centerline{\epsfxsize=5truein  \epsffile{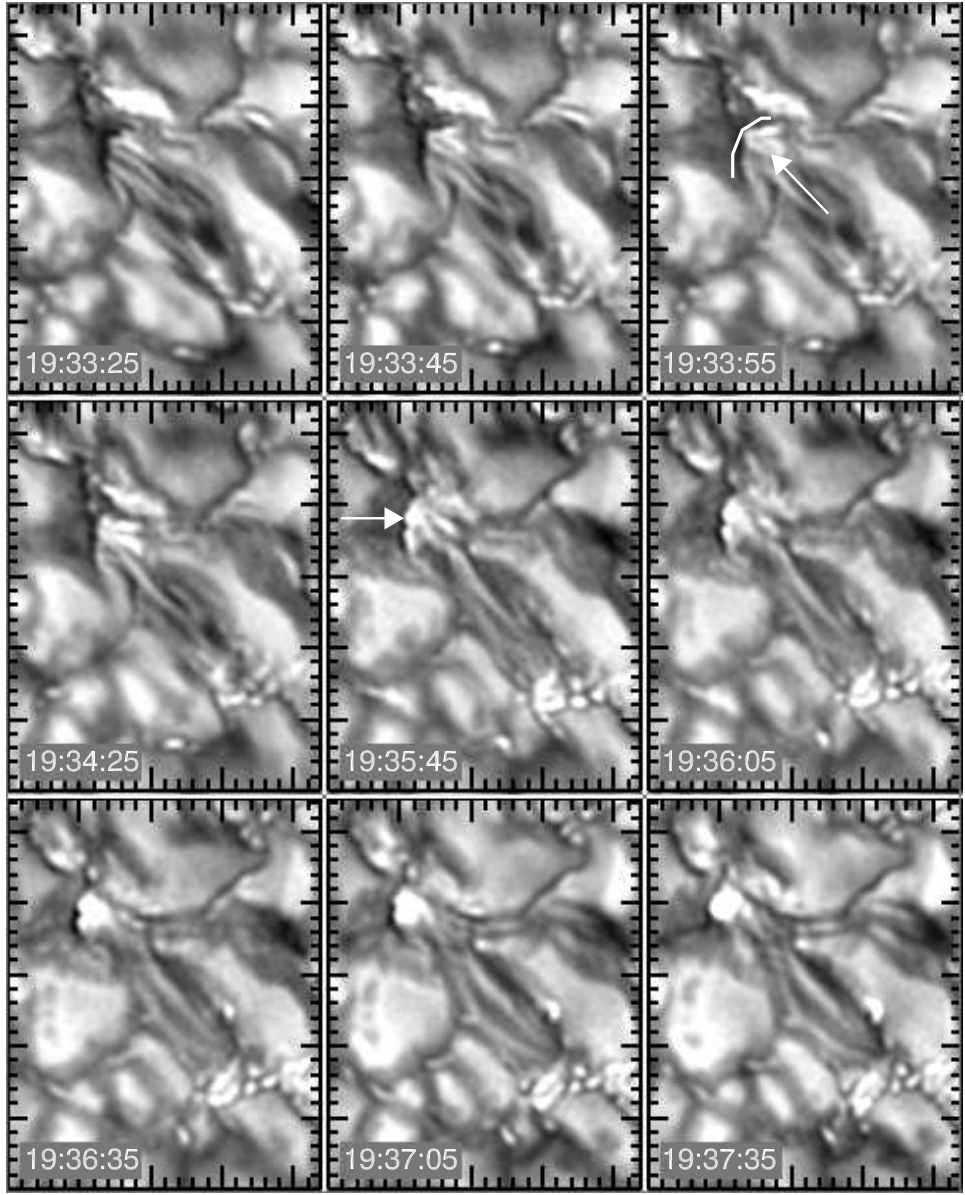}}
\caption{Further evolution of the TiO filamentary structure shown in Figure \ref{aug30}. The arrow in 19:33:55 panel indicates the direction of the motion of the filament's tip. The arc outlines the bright moving front that forms as the tip of the filament intrudes into the neighbouring intergranular lane. The arrow in the 19:35:45 panel indicates new bright points that formed as a results of the motion. Short tick marks separate 0.2~Mm spatial intervals.}
\label{bp_squeeze}
\end{figure}

Further evolution this filamentary structure between 19:33~UT and 19:37~UT is shown in Figure \ref{bp_squeeze}. The appearance of the HMI positive polarity was co-spatial and co-temporal with formation of bright points at the tip of the extending TiO filaments (arrow in the central panel). The data also show that the positive polarity was rapidly moving into the intergranular lane (arrow in 19:33:55 panel) creating a well defined moving front of enhanced brightness (arc in 19:33:55 panel). Such motions probably piled up and intensified magnetic fields as well as induced strong downflow lanes (see the dark lane on the left side of the bright ridge). The above suggests that these BPs could be due to convective collapse \citep{parker_bp_formation,spruit_bp_formation}. A similar event, associated with penumbral filaments, was recently described in \cite{Lim_2011}.

Based on visual inspection of various TiO data sets, we find that such events are quite frequent in active regions and plage areas, where the magnetic fields are relatively strong. To the contrary, they are quite scarce in quiet sun regions and coronal holes. In those rare instances when they do appear there, they tend to occur in association with network fields and/or clusters of bright points.

\subsection{Appearance of Bright Points inside a Granule}

\begin{figure}[t]
\centerline{\epsfxsize=4truein  \epsffile{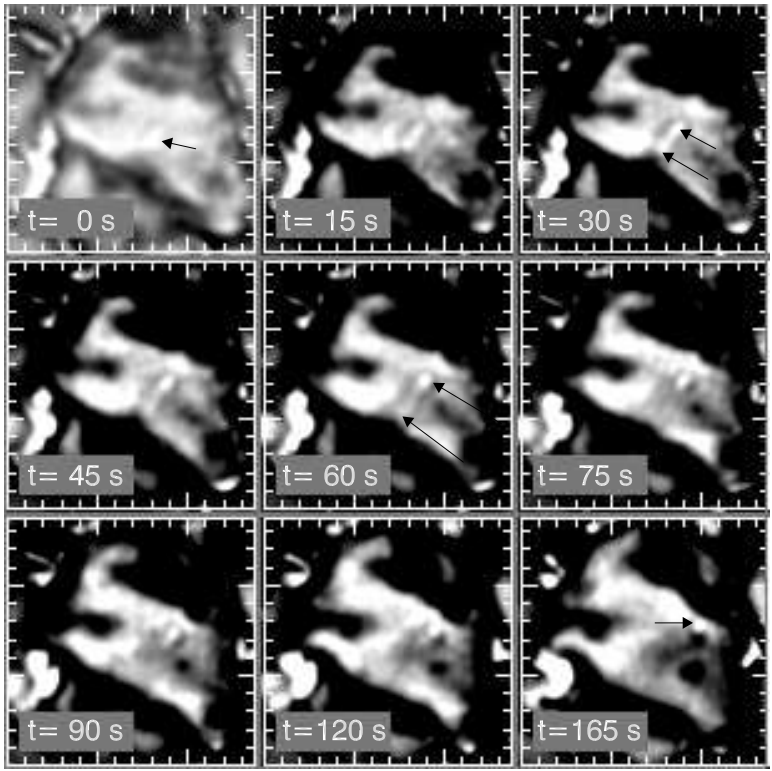}}
\caption{Sequence of TiO images showing appearance and evolution of bright point-like feature inside a granule. Time intervals for each image are shown in the lower left corner. Appearance of new flux is accompanied by development of a bright ridge inside the host granule. All panels but first were saturated in order to highlight the dimming surrounding the bright ridge. Short tick marks separate 0.2~Mm spatial intervals.}
\label{bpgr}
\end{figure}

\noindent In Figure \ref{bpgr} we show appearance of BP-like structures inside a granule. We do not have magnetic field data for this event and the following speculations are based on the similarity with the previous case. The event lasted for about 3~min and first appeared in the FOV as a 0.2~Mm roundish brightening situated in the middle of a granule (Figure \ref{bpgr}, arrow in t=0~s panel). The brightening grew into a ridge surrounded by a weak transient darkening (t=30~s panel). The two arrows in this panel indicate the extreme ends of the ridge. The t=45~s panel shows that one end (footpoint) became significantly brighter than the surrounding granule and shortly after one can clearly discern that the bright ridge has split into two bright points (at t=60~s), one of which shortly faded (at t=75~s). The remaining BP-like feature moved along a straight line toward the edge of the granule until it disappeared into the intregranular lane. The rate of footpoint separation, determined between t=15~s and t=60~s, was about 20~km s$^{-1}$. We speculate that the BP-like features are intensity counterparts of small-scale flux emergence inside a granule.

\section{Conclusion and Discussion}

\noindent The main inference of our study is that the new small-scale magnetic flux, passing through the photosphere, appears to leave different types of footprint on solar granulation: i) diffuse irregular patches of increased brightness evolving without any well defined scenario (pattern), ii) well defined filament-like structures that elongate as they evolve, and iii) formation of bright point-like features inside granules.

This is the first time we observe details of granular-scale flux emergence directly in the photospheric intensity images. To the best of our knowledge, these are also the first reported observations of appearance of bright point-like structures inside granules. Similarity with the filamentary flux emergence does suggest that the granular bright points are magnetic structures, however, spectra-polarimetry data with high temporal sampling (10~s) are needed to understand their nature.

The NST/IRIM magnetic field data also present some evidence that horizontal vortex tubes \citep{Steiner_2010} may to be associated with enhanced small-scale fields. These vortex tubes, easily detectable in photospheric intensity maps, were also reported to be at the origin of small intergranular jets \citep{goode_apjl_2010, Yurchyshyn_2011} first detected in NST off-band H$\alpha$ images.

We propose that the filamentary emergence events correspond to the transient darkenings observed earlier \citep[\textit{e.g.}][]{cheung_2008}. These events may also be related to the events recently described in \cite{Dominguez_2012}.  \cite{strous_zwaan_1999} first suggested that the transient darkenings correspond to crests of emerging magnetic flux since they usually have bright points at either ends and are associated with upflows of order of 0.5-1.0 km/s. However, it is rather difficult to reconcile the upflows with the transient darkenings that usually signal dense cool descending plasma \citep{cheung_2008,bharti_2011,rempel_2011}. The rising flux bundles, observed as bright filaments, may account for the blue-shifted Doppler signal detected by \cite{strous_zwaan_1999}. \cite{cheung_2008} suggested that a fast rising flux tube may push parcels of hot plasma into the upper photosphere, which overturn within a few pressure heights. The overturned plasma then drains down to the photosphere forming transient dark lanes around the rising bright filaments/flux tubes. Also, because of the small scale height, the rising flux tube expands rapidly when it leaves the photosphere. Strong lateral expansion and the downflows produce an impression in the granulation field seen as darkening.

\cite{ishikawa_2008} reported two types of flux emergence events. Type 1 events were argued to be buoyancy driven, while Type 2 events were thought to be convection driven flux emergence episodes. While both types have comparable magnetic field strength (600~G), size (about 1''.5) and the lifetime (about 4~min), they diverge in the values of magnetic filling factor (0.44 and 0.17 respectively). Type 1 events cause granule expansion we propose that they may correspond to the filamentary emergence discussed here. Although the orientation of these emerging bipoles appears random relative to the orientation of the associated active region, they tend to be oriented in such a manner that one of the footpoints faces and moves toward the neighbouring magnetic field concentration. This behaviour suggests that the filamentary emerging fields are deeper rooted and they are part of larger coherent structure.

The diffuse emergence may be understood assuming that it is convection-driven and governed by granular upflows. Their frequent appearance in quiet sun granulation hints that they could be driven by local processes such as field intensification by the granular scale local convection flows \citep[e.g.,][]{vogler_2008,schussler_2008,Steiner_2010}. \cite{Lites_2011} noted that only a very small correlation exists between magnetic flux imbalances determined for a subregion of very weak fields and the entire field of view (the large-scale structure). However, high correlation is expected in case the small-scale internetwork fields are purely due to shredding of the pre-existing fields of active regions, plages, etc. The absence of significant correlations favours the small-scale dynamo rather than field re-cycling. \cite{ishikawa_tsuneta_2010}, noted that the idea of small-scale fields to be a product of magnetic field recycling is not consistent with the fact that horizontal fields randomly appear on the solar surface in temporal domain.

Authors thank BBSO observers and the  engineering team for their contribution to this study. Authors thank M. Cheung and M. Rempel as well as the ISSI International Team lead by I.N. Kitiashvili at ISSI (International Space Science  Institute) in Bern for insightful discussions. VY work was partly supported under NASA GI NNX08AJ20G and LWS TR\&T NNG0-5GN34G grants. VA acknowledges partial support from NSF grant ATM-0716512. PG, VA and VY are partially supported by NSF (AGS-0745744) and NASA (NNY 08BA22G). PG is partially supported by AFOSR (FA9550-09-1-0655).

\

\end{document}